\documentclass{IEEEcsmag}
\usepackage{url}
\usepackage{pifont}
\usepackage[many]{tcolorbox}
\let\labelindent\relax
\usepackage{enumitem}
\usepackage{tikz}
\usepackage{textcomp}
\newcommand{\staree}{\ding{72}}
\newcommand{\snowflake}{\ding{100}}
\newcommand{\trianglee}{\ding{115}}
\newcommand{\checkee}{\ding{51}}
\newcommand*\circled[1]{\tikz[baseline=(char.base)]{
		\node[shape=circle,draw,inner sep=2pt] (char) {#1};}}
\definecolor{my-blue}{cmyk}{0.6, 0.31, 0, 0.5}
\NewTColorBox[blend into=tables]{NewBox}{ s O{!htbp} }{
	floatplacement={#2},
	IfBooleanTF={#1}{float*,width=\textwidth}{float},
	colframe=blue!50!black,colback=blue!10!white,
	breakable,
	enhanced,
	label=tab:paperF:researchMethod,
	title=Context of this case study.,
	segmentation style={solid, my-blue!30},
	colback=white,
	colframe=my-blue,
	arc=0.9em,
	bottom= 0pt,
	top =5pt,
	boxsep=0.5mm,
}
\usepackage[colorlinks,urlcolor=blue,linkcolor=blue,citecolor=blue]{hyperref}

\jvol{XX}
\jnum{XX}
\paper{}
\jmonth{}
\jname{}
\pubyear{}

\begin{document}

	\title{Why and How Your Traceability Should Evolve: Insights from an Automotive Supplier}
	\author{Rebekka Wohlrab}
	\affil{Systemite AB, Gothenburg, Sweden and Chalmers $|$ University of Gothenburg, Sweden}
	\author{Patrizio Pelliccione}
	\affil{University of L'Aquila, Italy and Chalmers $|$ University of Gothenburg, Sweden}
	\author{Ali Shahrokni}
	\affil{Zenuity AB, Gothenburg, Sweden}		
	\author{Eric Knauss}
	\affil{Chalmers $|$ University of Gothenburg, Sweden}

	\begin{abstract}
	Traceability is a key enabler of various activities in automotive software and systems engineering and required by several standards.
	However, most existing traceability management approaches do not consider that traceability is situated in constantly changing development contexts involving multiple stakeholders.
	Together with an automotive supplier, we analyzed how technology, business, and organizational factors raise the need for flexible traceability.
	We present how traceability can be evolved in the development lifecycle, from early elicitation of traceability needs to the implementation of mature traceability strategies.
	Moreover, we shed light on how traceability can be managed flexibly within an agile team and more formally when crossing team borders and organizational borders.
	Based on these insights, we present requirements for flexible tool solutions, supporting varying levels of data quality, change propagation, versioning, and organizational traceability. \newline
	
	
	\end{abstract}
	\maketitle

	\chapterinitial{Traceability} is a recognized necessity for practitioners aiming to comply with safety or quality standards, to track development progress, or to analyze the impact of change.
		\begin{tikzpicture}[overlay]
	\node[draw, fill=white, thick] (b) at (0,20.5){		\sffamily\textcopyright  2020 IEEE --- Accepted for publication in IEEE Software};
	
	\end{tikzpicture}
	A variety of approaches for various traceability-related activities exist, ranging from the definition of a traceability strategy to the creation, maintenance, and use of trace links.
	A \emph{traceability strategy} defines a traceability information model (TIM), processes, and tooling~\cite{Gotel2012Fundamentals}.
	Some empirical findings indicate that one should define traceability strategies for projects upfront~\cite{Rempel2013}.
	Others state that traceability strategies need to be introduced incrementally and adjusted, but it is unclear how these adjustments should be performed in practice~\cite{Neumuller2006}.
	When setting up traceability strategies and processes, stakeholders' needs are different than when data becomes more mature and development artifacts are changed and updated, rather than written from scratch.
	Many approaches in the literature neglect that needs for traceability change over time and depend on the stakeholders that create, maintain, and use traceability.
	
	In this article, we focus on ``\textit{ingredients for through-life traceability success}''~\cite{Cleland-Huang2014}, one of the research directions related to traceability strategizing.
	Our contributions relate to high-level activities of the traceability lifecycle~\cite{Gotel2012Fundamentals,Cleland-Huang2014} and present empirical insights into traceability strategizing gathered over the course of three years at an automotive supplier.
	While it is important to evolve trace links with suitable approaches~\cite{Rahimi2018}, also the high-level traceability strategy and information models need to be evolved.
	
	In a previous study, we investigated challenges and opportunities with collaborative traceability management~\cite{Wohlrab2018REEN}.
	In practice, trace links between artifacts created in multiple teams have different characteristics than intra-team trace links.
	We focus on an automotive supplier to analyze how the heterogeneity of trace links in terms of time and organizational settings can be better supported by methods and tools.
	Traceability in automotive contexts is especially challenging:
	automotive systems need to fulfill multiple quality attributes, have heterogeneous functions, and are developed over several years by multidisciplinary stakeholders located at suppliers and OEMs~\cite{Ebert2017}.
	We describe four types of trace links depending on involved organizational groups, from trace links used within an agile team to links crossing company borders.
	We examine how changes to the traceability strategy have been performed over time and present the OADI cycle of traceability.
	Finally, we describe requirements for tooling that supports varying levels of data quality, change propagation, versioning, and organizational traceability.

	\section{Background}
	Traceability is concerned with the creation and use of trace links.
	A \textit{trace link} is a connection from one or more source artifacts to one or more target artifacts.
	Trace links can have types, attributes, and versions.
	Artifacts can be of different types (e.g., code function, test case, or requirement) and be represented in text, models, tables, etc.
	When studying artifacts in systems engineering, distinguishing between \emph{boundary objects} and \emph{locally relevant artifacts} is often useful~\cite{Wohlrab2019JSME}.
	Boundary objects are objects that can be used to cross team borders in large-scale development and create a common understanding, while preserving each team's identity~\cite{Wohlrab2019JSME,Star1989}.
	Locally relevant artifacts are created, maintained, and used by members of one agile team.
	
	Typical boundary objects include interfaces, requirements, or architecture descriptions.
	In practice, stakeholders often work on the areas of function definition, design, implementation, and testing.
	\textit{Vertical boundary objects}~\cite{Wohlrab2019JSME} are used by groups on different levels of abstraction and help to create a common understanding between these areas, e.g., in the form of functional requirements used by requirements engineers, safety experts, and testers.
	\textit{Horizontal boundary objects}~\cite{Wohlrab2019JSME}, on the other hand, are used by groups on the same level of abstraction.
	Examples are signals between components developed by two agile teams.

\section{Traceability in Practice: An Example} \label{sec:runningexample}
\begin{figure*}
	\centering
	\includegraphics[width=\linewidth]{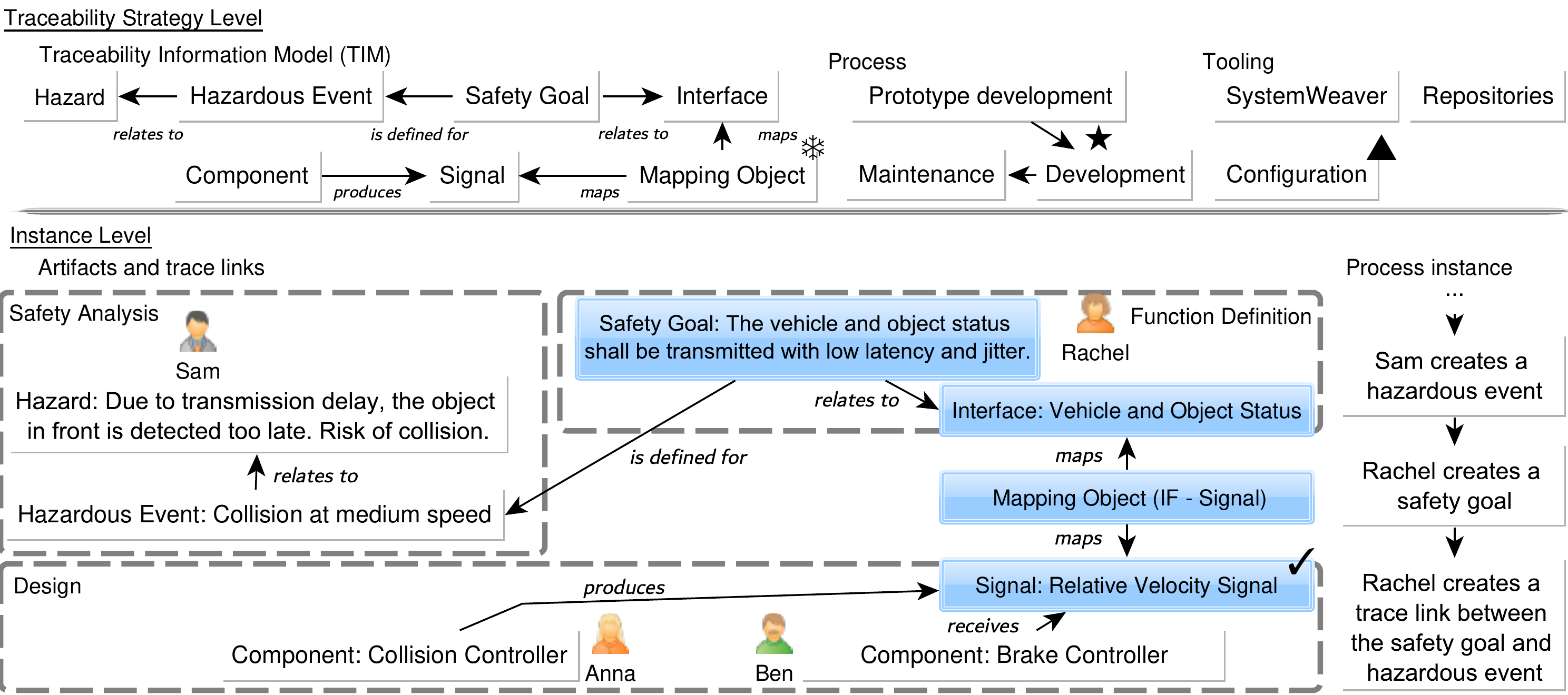}
	\caption{Running example related to the development of a collision avoidance by braking system. The top part shows the traceability strategy level (TIM, process, and tooling). The lower part shows the instance level: Artifacts, trace links, and a process instance. Boxes with dashed lines represent three areas (function definition, safety analysis, and design) by which artifacts are grouped. Arrows represent trace links and are labeled with trace link types. Boundary objects have a blue and locally relevant artifacts have a white background.}
	\label{fig:paperF:runningexample}
\end{figure*}

We consider an example related to a \textit{collision avoidance by braking} system~\cite{Bond2003}, shown in Figure~\ref{fig:paperF:runningexample}.
The context of the case study and the top part of Figure~\ref{fig:paperF:runningexample} are described in Table~\ref{tab:paperF:researchMethod}.
The lower part of the figure represents the concrete artifacts and trace links, as well as a process instance.
The safety analyst Sam identifies a potential hazard: an object is detected late because of a transmission delay.
A related hazardous event is a collision at medium speed.
The function expert Rachel defines a safety goal concerned with latency and jitter.
Rachel traces it to a function interface: \textit{Vehicle and object status}.
The function interface is linked to a \textit{relative velocity signal} on the design level.
To support decoupling, the linking is done using a \textit{mapping object}, i.e., a separate artifact with trace links to two artifacts that shall be related.
Anna's \textit{collision controller} component detects an object and calculates its velocity relative to the vehicle.
It produces a signal which is processed and traced to by Ben's \textit{brake controller} to potentially initiate braking.

\begin{NewBox}*
	We conducted this study with Zenuity, an automotive supplier developing highly critical software for several automotive OEMs.
	Zenuity has been using agile methods for 3~years and is organized in agile teams with 6--8 members.
	There exist 60--70 component teams, many of which also have responsibilities for a feature.
	Approximately 2--3 teams are pure feature teams.
	
	One of the authors of the article is employed at Zenuity and another one at Systemite, the provider of the tool SystemWeaver.
	The industrial authors have followed the development of traceability management practices over the course of three years, taken meeting notes, and recorded changes to the traceability strategy over time.
	Additionally, we conducted six semi-structured interviews with four process, methods, and tool experts, one safety expert and function owner, and one developer.
	The interviews were analyzed with an open coding approach and relations between codes were discussed to arrive at our findings.
	Besides, we were involved in several research initiatives in a long-term engagement.
	In Vinnova FFI projects on the Next Generation Electronic Architecture in automotive companies\footnote{\url{https://www.vinnova.se/p/next-generation-electrical-architecture/}}, we investigated how a new electronic architecture can facilitate cross-organizational collaboration.
	Moreover, the Software Center\footnote{\url{https://www.software-center.se/}} Project 27 on RE for Large-Scale Agile System Development allowed us to collaborate with more than ten companies on traceability-related topics.
	We triangulated Zenuity's experience with these companies.
	\tcbline
	
	The top part of Figure~\ref{fig:paperF:runningexample} shows an excerpt of Zenuity's traceability strategy-level information with a TIM, process, and tooling as core elements of a traceability solution.
	The TIM shows the key artifact types and trace link types.
	The process shows the abstract activities \textit{prototype development}, \textit{development}, and \textit{maintenance}, for which subprocesses are defined.
	Repositories are used to version code in files (e.g., C++ or Matlab).
	Besides using repositories, information is stored in the systems engineering tool SystemWeaver\footnote{\url{https://systemweaver.se/}} to define systems' functionality, design, safety analysis, and testing.
	In this paper, we will exemplify how the traceability strategy has evolved, indicated by symbols in the figure (\snowflake, \staree, \trianglee, \checkee).
	When leaving the prototype development stage (\staree), a mapping object was introduced (\snowflake), the configuration adjusted (\trianglee), and a signal became a boundary object (\checkee).
	
	SystemWeaver is a highly configurable, collaborative modeling platform with a TIM that can be tailored to the customers' needs.
	Both forward and backward traceability are supported and trace links on various levels of granularity can be established.
	Trace links can be direct or implemented using a ``\textit{mapping object}'' that refers to two decoupled artifacts, as we illustrated in the example.
	SystemWeaver includes support for versioning and quality checks.
	Artifacts and trace links can have a status specifying whether the artifact is ``\textit{in work}'', i.e., not yet stable, or at a particular version.
	It also supports visualization and representation of data, e.g., in tables, reports, graphs, charts, or export formats that can be configured by the end users.
	For instance, requirements are typically exchanged between companies by exporting and importing files that comply to the ReqIF format\footnote{\url{https://www.omg.org/spec/ReqIF/About-ReqIF/}}.
	For each artifact, one can see who created and changed it at what point in time and who is the responsible owner.
	Traceability can be established using typed links between artifacts, with all types defined in the TIM.
	Besides typed links with strong semantics, it is also possible to create generic hyperlinks between artifacts in SystemWeaver.
\end{NewBox}

\section{Organizational Settings Influence Coordination Mechanisms Around Traceability}
\begin{figure}[b]
	\centering
	\includegraphics[width=\linewidth]{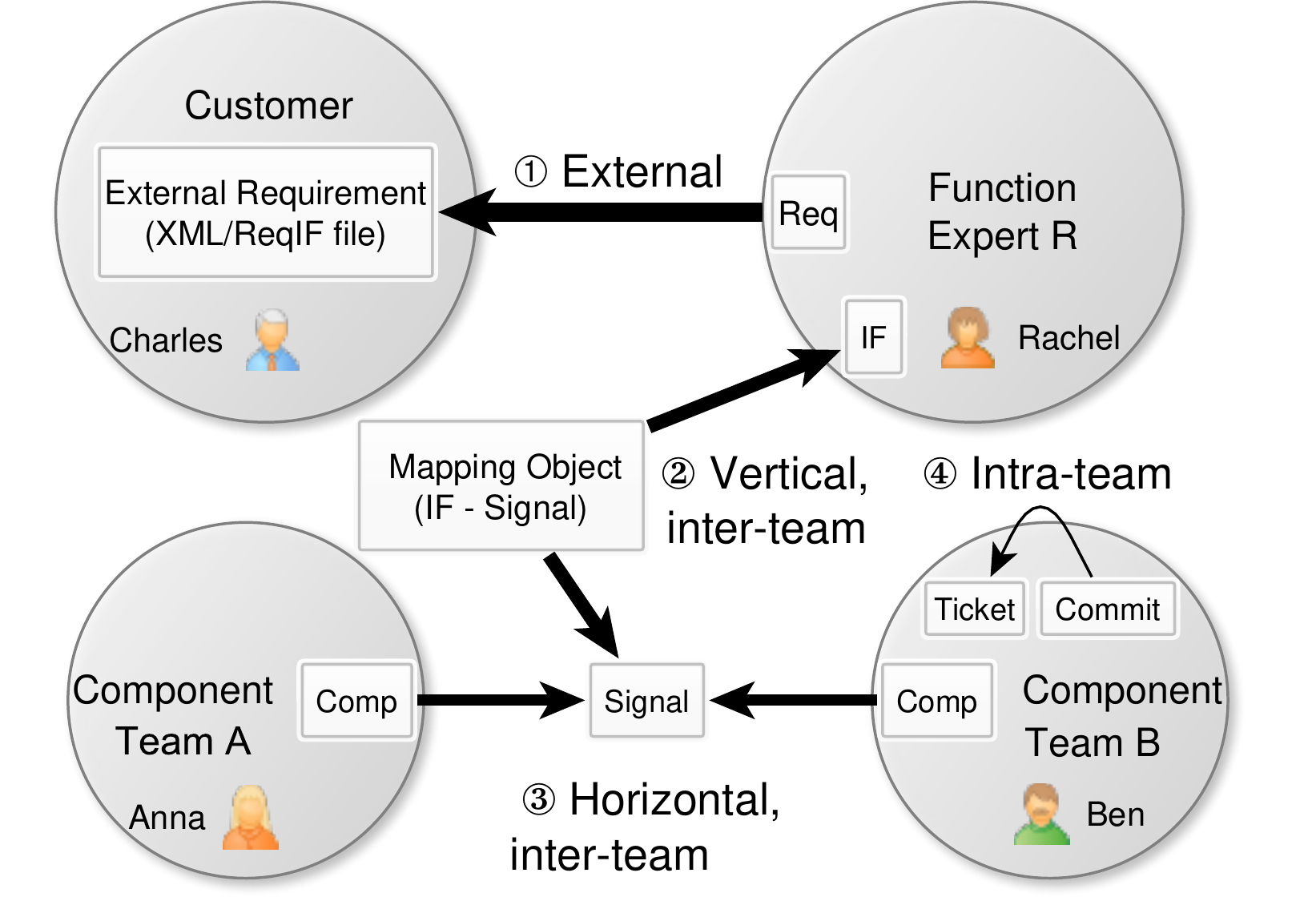}
	\caption{Trace links between artifacts (white boxes) used by organizational groups (gray circles)}
	\label{fig:paperF:orga}
\end{figure}

Organizational settings greatly impact coordination mechanisms around traceability.
Figure~\ref{fig:paperF:orga} shows stakeholders, their organizational groups, and artifacts that they manage.
Component teams operate at the same level of abstraction, whereas others bridge an abstraction gap (e.g., when tracing from abstract function interfaces to signals).
Trace links between artifacts are shown as black arrows.
The thicker the arrows, the more rigid are the trace links.
Rigidity relates to how strictly trace links are documented and coordinated.
Rigid trace links undergo formal coordination, versioning, and change management.
Four groups of links are shown in Figure~\ref{fig:paperF:orga} with labels \circled{i}.

At Zenuity, ReqIF files with external requirements are imported and traced \circled{1} from a function expert's functional requirements.
These trace links are important to show that all customer requirements have been considered.
The intention is to avoid changes and keep external links stable.
All artifacts have IDs and version numbers to improve traceability across tool borders.
All changes are rigorously documented and formally discussed in meetings with meeting minutes.

Group \circled{2} links are used across different levels of abstraction.
In our example, function interfaces serve as vertical boundary objects between the function expert and component teams and are linked to signals using mapping objects.
Mapping objects establish a connection but allow artifacts to be refined independently from each other.
The links between function artifacts and design artifacts are important to keep track of which functions are implemented in which design components.
In an agile environment, interfaces and signals are versioned and evolve further.
Change can happen in a top-down manner, but also bottom-up: when the concrete artifacts change, the more abstract ones are updated as well.
In this sense, vertical boundary objects are living artifacts that support change impact analysis and are continuously adjusted based on stakeholders' needs.

Group \circled{3} links affect teams on the same level of abstraction, and close in their disciplines, as well as geographically and temporally~\cite{Bjarnason2017}.
In the example, signals are used as horizontal boundary objects between component teams.
They need to be consistently used so that the developed system can operate correctly.
Component teams coordinate changes through meetings, workshops, or an issue tracking system.
Especially in early phases, signals are not formally versioned, but quickly changed and coordinated in meetings.
With time, a signal between two components might become relevant for other components and therefore require stricter coordination mechanisms.

Group \circled{4} links are used within a team, e.g., to trace internal documentation, local design decisions, or commit messages.
If possible, one aims to reduce the effort of versioning and change management, but aims for lightweight coordination in face-to-face meetings.
Sometimes, restructuring of the architecture can imply that a signal governed by a team becomes relevant for other components and teams.
As this signal becomes a boundary object, needs for change management, data quality, and versioning might change.

Our observations show that cost-efficient traceability management requires situation-specific coordination mechanisms.
Face-to-face communication is used within agile teams and formal alignment, versioning, and data quality standards are required when communicating across team borders.
To support change, boundary objects create a common understanding and need to fulfill certain properties to stay useful and consistent.
They cannot exist in isolation, but need to be traceable from artifacts by different teams.
We use the term ``\textit{living boundary objects}'' for boundary objects that are traced to other artifacts, kept up to date, and serve for inter-team coordination.
For instance, the safety goal in Figure~\ref{fig:paperF:runningexample} can serve as a living boundary object because Sam, Rachel, Anna, and Ben have meetings in which they discuss concrete latency values, keep artifacts up to date, and maintain traceability.
Living boundary objects ensure that a common understanding is maintained---even during the evolution of traceability strategies.

\section{How to Develop Traceability Over Time}
\begin{figure}
	\centering
	\includegraphics[width=\linewidth]{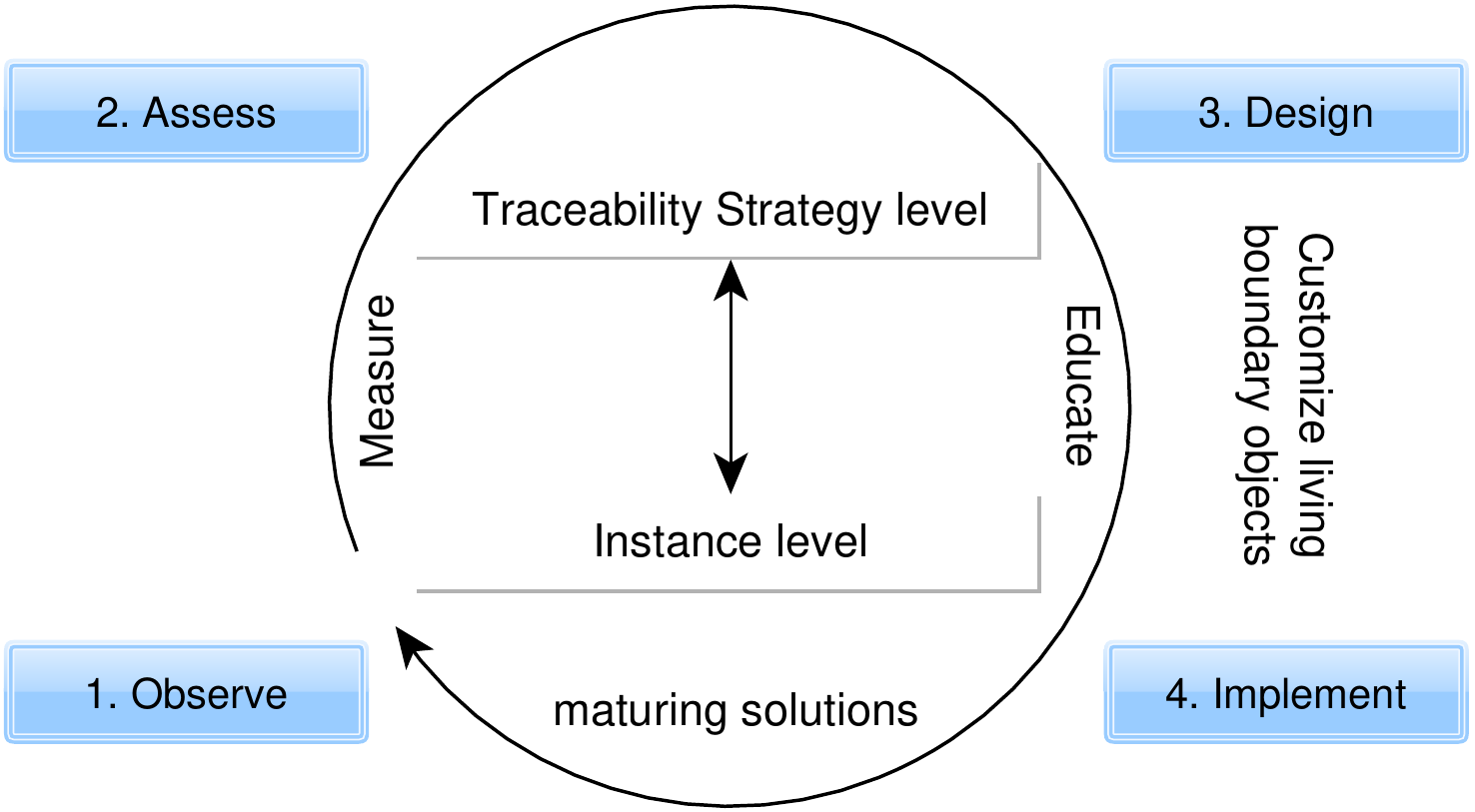}
	\caption{OADI cycle of traceability (inspired by Kofman, referenced in~\cite{Kim1998})}
	\label{fig:paperF:changeovertime}
\end{figure}

The process of establishing and tailoring traceability in an organization can be explained with learning models.
We found the OADI (Observe-Assess-Design-Implement)~\cite{Kim1998} cycle applicable to traceability.
Humans \textit{observe} the environment, \textit{assess} it by reflecting on observations, \textit{design} concepts to meet practical needs, and eventually \textit{implement} the design.

Zenuity created common traceability practices over the course of three years.
Figure~\ref{fig:paperF:changeovertime} depicts the OADI cycle of traceability.
Traceability practices were iteratively created and instantiated.
Iterations had varying lengths, from two weeks during the initial phases to six months.
We found substantial differences in the first iterations of the OADI cycle in comparison to when traceability had become more established.

\paragraph{Phase 1: Observe}
In the first iterations of the OADI cycle, rather than relying on instantiated data, meetings and informal discussions were used to observe how stakeholders planned to work and what traceability needs they had.
At Zenuity, experts elicited important product aspects, the roles of external customers, and the creation of the organizational structure and processes.
Business needs and technology trends were observed as they might affect the traceability strategy.

In later iterations, observations were continuously supported by more than 40 quality checks at Zenuity.
Completeness was visualized by a pie chart indicating how many external requirements were traced to functional requirements.
Other checks related to validity (``\textit{is the instance data in line with the TIM?}''), and version consistency.
Data quality is not only concerned with trace links, but also with connected artifacts.
Four properties of data quality are checked:
\begin{itemize}
	\item {\em Consistency} refers to the absence of differences between artifacts representing the same concepts.
	For instance, if a function interface uses the type Integer, the corresponding signal should use the same type.
	\item \textit{Version consistency} is concerned with the version numbers of artifacts and trace links. It is typically violated if trace links to multiple versions of the same artifact exist within a baseline.
	\item {\em Validity} is a measure of conformity of an artifact to another artifact or established rules and templates.
	When changing the TIM, instance data needs to be adjusted to ensure validity.
	\item {\em Completeness} refers to an indication of the comprehensiveness of data to address specific needs.
	At Zenuity, all signals defined for a system should be used as in-~and output signals of at least one component. 
\end{itemize}

\paragraph{Phase 2: Assess}
Based on observations, Zenuity reassessed the tool configuration at certain delivery dates and the end of every sprint (\trianglee).
When an initial TIM had been defined, stakeholders checked which parts of the TIM were instantiated as intended and which parts end users struggled with.
Acceptance criteria were tailored to support the assessment.
While a safety analysis was not mandatory for acceptance in the first months, a complete analysis for each function was soon included in the acceptance criteria.

The traceability strategy had to be adjusted, e.g., after the prototype stage (\staree).
After one year, the architecture was restructured during several sprints, which influenced interfaces between components.
A signal that was relevant within a team became a boundary object and used by several teams (\checkee).
Stakeholders assessed coordination mechanisms for artifacts and links in every sprint and changed them several times a year, e.g., when new customers were added and new boundary objects were created.

\paragraph{Phase 3: Design}
Whenever necessary, solutions were designed based on the previous assessment.
For example, decoupling was more important at certain stages.
After three months, methods and tools experts adjusted the TIM and introduced mapping objects to decouple interfaces and signals (\textbf{\snowflake}).
Moreover, technology changes raised the need for new types in the TIM.
Figure~\ref{fig:paperF:runningexample}'s structure with areas for function definition, safety analysis, and design was designed.
Relying on boundary objects helped to support change and maintain a common understanding across sites.
Stakeholders conceived living boundary objects and guidelines to trace artifacts, keep them up to date, and meet quality goals.

\paragraph{Phase 4: Implement}
Stakeholders implemented design concepts by changing the TIM, adjusting quality checks, or refining training material to communicate guidelines.
Typically, users were trained in four-hour workshops, considering role-specific perspectives of safety experts, developers, or function experts.
Material was made accessible through SystemWeaver.
The culture needed to be fostered and traceability management established as a natural part of teams' work.
In every sprint, traceability-related DoD criteria were checked (e.g., regarding completeness).

In early iterations, several parts of the TIM were created simultaneously and changed in every sprint, whereas after 12 months, only smaller changes were performed, e.g., refining the names of trace link types or facilitating integration with other tools.
During prototype development, teams explored solutions that were not traced to boundary objects yet.
The set of trace links was not complete and consistent.
Prototype development involved creating loosely coupled information to develop ideas.
After one year, ideas had matured and were refined to meet quality needs,
traced to boundary objects, and became more formal.
The increase in formality was supported by quality checks and dedicated meetings.
Formal versioning was established, especially when external organizations were involved.
Versioning gave stakeholders more control of when and how changes happened.
To indicate the severity of quality issues, levels of warnings were created and adjusted over time.
Defined semantics of link types became more important.
An architect stated that eventually, a ``\textit{point of maturity}'' was reached.
At that point, changes became less common, until technology, business, process, or organizational changes motivated the need to evolve traceability further.

\section{Requirements for More Flexible Traceability Management}
We have presented how Zenuity's trace links were managed, depending on the involved organizational groups and the current phase.
For instance, we described how technology and business changes impacted the evolution of the TIM and traceability strategy (e.g., when new signals emerged as boundary objects or when the TIM was changed to improve decoupling of artifacts).

SystemWeaver supports the required flexibility, e.g., by providing a configurable TIM designed to meet stakeholders' needs, quality checks, guidelines for versioning, or conventions to manage change.
Traceability is mostly managed manually and more than 200,000 trace links are in use.
To support flexible traceability even better in future tools, we identified the following requirements.
\paragraph{1. Data Quality Needs} \label{req:DataQuality}
As stated in previous sections, data quality is essential.
However, data quality needs change over time and assessment criteria should be tailored to stakeholders' needs.
For instance, incomplete safety analysis can be accepted early on, but not when delivering a product to customers.
In prototype phases, inconsistent and incomplete trace links and artifacts are acceptable, whereas closer to a release, data quality is enforced.
When Zenuity replaced direct trace links with mapping objects to decouple interfaces and signals in the TIM, invalid trace links were temporarily accepted, before the instance data had been adjusted.
On the traceability strategy level, a tool could exploit historical data to provide suggestions for data quality needs in a specific phase.
For instance, more checks might be activated when a signal becomes a boundary object between teams.
\paragraph{2. Type of Versioning}
A tool could suggest more fine-grained versioning mechanisms, e.g., loose, mild, and strict versioning.
\textit{Loose versioning} (in early phases) could allow users to create new versions whenever needed, but changes do not automatically result in new versions.
\textit{Mild versioning} of a locally relevant artifact could create intermediate versions whenever a change to the artifact is related to a major change of a boundary object, but rely on loose versioning otherwise.
\textit{Strict versioning} of fine-grained deltas could be established in later phases.
Generally, links to boundary objects should likely be versioned early on, while intra-team links can stay loosely versioned during certain iterations of the OADI cycle.
\paragraph{3. Organizational Traceability}
For each artifact, responsible users should be captured along with roles and relations to other users.
Whenever trace links or artifacts are changed, the tool should support updating organizational traceability in a (semi-)automatic way.
As an artifact becomes relevant to create a common understanding across team borders, a particular stakeholder could become responsible for that boundary object.
On a strategy level, it could be defined what organizational aspects should be captured and to what extent users should be involved (e.g., when updating lists of users that are affected by a boundary object).
Visualization mechanisms and notification features~\cite{Wohlrab2018REEN} can support the collaborative management of organizational traceability.
\paragraph{4. Change Management Support} \label{req:userChangeManagement}
Changes to artifacts and trace links need to be managed over time.
For instance, moving a signal to the scope of an individual team changes the desired quality needs and coordination mechanisms.
Change management should be supported for the evolution of trace links and artifacts~(e.g., \cite{Rahimi2018}) and also on the strategy level.
Based on the needs in particular phases and the living boundary objects in use, levels of rigidity, quality, and versioning should change.
Tools can support these concerns:
\begin{enumerate}[label=\alph*)]
	\item {\em Detection of change:} A detector should monitor the access to artifacts by users and the potential impact of changes on other artifacts.
	Historical information about co-changes can help to identify what artifacts and trace links are likely affected by a change, e.g., using Bayesian networks~\cite{Mirarab2007} or other machine learning techniques.
	Artifacts that likely affect multiple teams are candidates for boundary objects~\cite{Wohlrab2019JSME}.
	\item {\em Change propagation and model evolution:}
	Affected artifacts and trace links need to be evolved~\cite{Rahimi2018}.
	If an artifact is detected as a candidate for a living boundary object, organizational traceability should be updated and a responsible owner of the boundary object might be semi-automatically identified.
	Moreover, a tool could suggest creating a Community of Practice or strict versioning for artifacts.
	If the TIM has been changed in a certain iteration, instance data should be evolved.
	For instance, if a link type via a mapping object is changed to a direct trace link type, a tool should update affected links.
	\item {\em Trigger tuning:} Tools should help to tailor triggers for change detection and quality.
	In initial phases, completeness can be neglected, but close to a release, all central artifacts and links should be in place.
	Similarly, versioning mechanisms should be adjusted.
	When an initial design is explored, loose versioning can be beneficial, whereas close to a release strict versioning is required.
\end{enumerate}

\section{Conclusion}
People and organizational distance between them impact how often and rigidly trace links are changed: lightweight and informal approaches are required for intra-team links, whereas more rigid and formal mechanisms are needed for larger organizational distance.
We presented how traceability has been established at an automotive supplier from the initial observation of stakeholders' needs to phases of high quality and formal versioning, as well as phases in which technology or business changes raised the need for adjustments.
Different lifecycle phases and involved organizational groups call for evolving approaches on the traceability strategy level.
Companies do not need to design the perfect traceability strategy from the start, but can adjust needs over time.
For instance, consistency was less important in early development, but crucial right before a release.
Change is and will be inevitable in large-scale organizations, especially as we move to more inter-company agility in the future.
Technology or business changes can motivate the need to extend the TIM or adjust quality needs.
To react to changes, organizations can benefit from living boundary objects that are traced to other artifacts, kept up to date, and serve for inter-team coordination.
We concluded that flexible tools need to support varying levels of data quality, versioning, organizational traceability, and change management.
In the automotive domain, but also in less challenging domains, practitioners can use our findings to create traceability strategies that are tailored to changing needs over time.
Companies and researchers can build upon our findings to improve solutions that provide adjustable support for the creation, change, quality assurance, and versioning of trace links and living boundary objects.

\section{Acknowledgment}
This work was partially supported by the Software Center Project 27 on RE for Large-Scale Agile System Development, the Wallenberg AI, Autonomous Systems and Software Program (WASP) funded by the Knut and Alice Wallenberg Foundation, and the Centre of EXcellence on Connected, Geo-Localized and Cybersecure Vehicles (EX-Emerge), funded by Italian Government under CIPE resolution n. 70/2017 (Aug. 7, 2017).

\begin{IEEEbiography}{Rebekka Wohlrab} is with Chalmers University of Technology and Systemite AB, Gothenburg, Sweden. 
	\begin{wrapfigure}{l}{1in}
		\vspace{-1.5em}
		\begin{center}
			\includegraphics[width=1in,height=1.25in,clip,keepaspectratio]{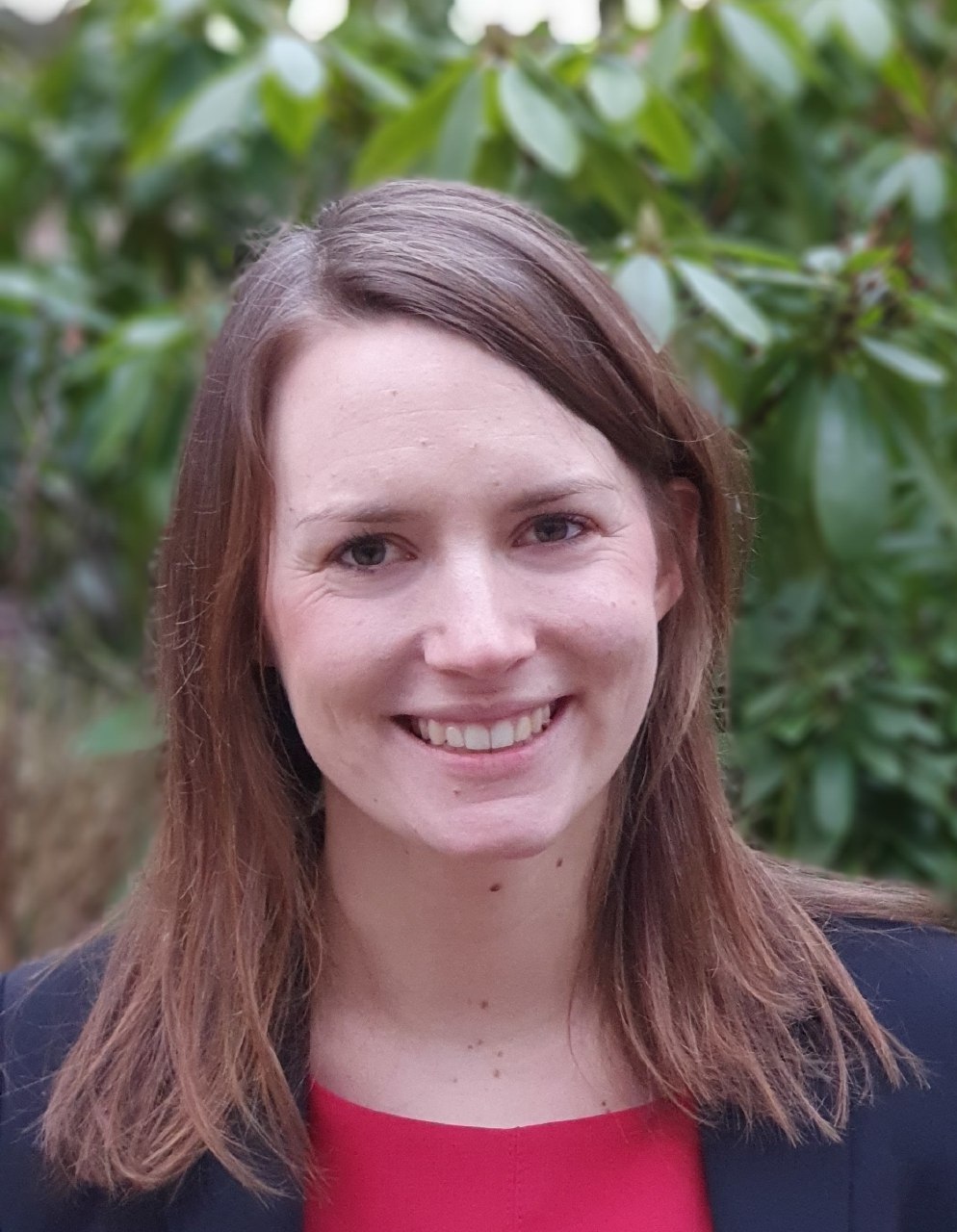}
		\end{center}
	\end{wrapfigure}
	Her research interests focus on software architecture and knowledge management in large-scale agile development.
	She holds a PhD from Chalmers University of Technology.
	Contact her at wohlrab@chalmers.se.
\end{IEEEbiography}

\begin{IEEEbiography}{Patrizio Pelliccione}
	is Associate Professor at the
	University of L'Aquila, Italy, and 
	\begin{wrapfigure}{l}{1in}
		\begin{center}
			\includegraphics[width=1in,height=1.25in,clip,keepaspectratio]{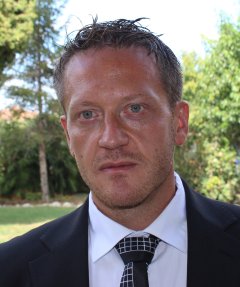}
		\end{center}
	\end{wrapfigure}		
	Chalmers $|$ University of Gothenburg, Sweden.
	His research topics are mainly in software architectures modeling and verification, autonomous systems, and formal methods.
	He holds a PhD from the University of L'Aquila.
	He has been on the program committees for top conferences and a reviewer for top journals on software engineering.
	Contact him at patrizio.pelliccione@univaq.it.
\end{IEEEbiography}

\begin{IEEEbiography}{Ali Shahrokni}
is with Zenuity AB, Gothenburg, 
\begin{wrapfigure}{l}{1in}
	\begin{center}
		\includegraphics[width=1in,height=1.25in,clip,keepaspectratio]{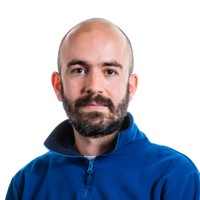}
	\end{center}
\end{wrapfigure}	
	Sweden.
	In his current position, he analyzes data flows and processes to create solutions for more efficient data and process management on an organizational level.
	He received a PhD degree at Chalmers University of Technology, Sweden.
	Contact him at ali.shahrokni@gmail.com.
\end{IEEEbiography}

\begin{IEEEbiography}{Eric Knauss}
	is Associate Professor at 
\begin{wrapfigure}{l}{1in}
	\begin{center}
		\includegraphics[width=1in,height=1.25in,clip,keepaspectratio]{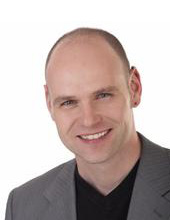}
	\end{center}
\end{wrapfigure}
	Chalmers $|$ University of Gothenburg, Sweden.
	His research interest focuses on managing requirements and related knowledge in large-scale and distributed software projects.
	He holds a PhD from Leibniz Universit\"{a}t Hannover, Germany.
	He is member of program and organization committees of top conferences and reviewer for top journals.
	Contact him at eric.knauss@cse.gu.se.
\end{IEEEbiography}


\begin{thebibliography}{10}
	
	\bibitem{Gotel2012Fundamentals}
	O.~Gotel, J.~Cleland-Huang, J.~{Huffman Hayes}, A.~Zisman, A.~Egyed,
	P.~Gr{\"{u}}nbacher, A.~Dekhtyar, G.~Antoniol, J.~Maletic, and
	P.~M{\"{a}}der, ``Traceability fundamentals,'' in {\em Software and Systems
		Traceability} (J.~Cleland-Huang, O.~Gotel, and A.~Zisman, eds.), pp.~3--22,
	Springer London, 2012.
	
	\bibitem{Rempel2013}
	P.~Rempel, P.~M{\"a}der, and T.~Kuschke, ``An empirical study on
	project-specific traceability strategies,'' in {\em RE'13}, pp.~195--204,
	2013.
	
	\bibitem{Neumuller2006}
	C.~{Neum\"{u}ller} and P.~{Gr\"{u}nbacher}, ``Automating software traceability
	in very small companies: A case study and lessons learned,'' in {\em ASE'06},
	pp.~145--156, Sep. 2006.
	
	\bibitem{Cleland-Huang2014}
	J.~Cleland-Huang, O.~C.~Z. Gotel, J.~Huffman~Hayes, P.~M\"{a}der, and
	A.~Zisman, ``Software traceability: Trends and future directions,'' in {\em
		FoSE'14}, pp.~55--69, 2014.
	
	\bibitem{Rahimi2018}
	M.~Rahimi and J.~Cleland-Huang, ``Evolving software trace links between
	requirements and source code,'' {\em Empirical Software Engineering},
	vol.~23, no.~4, pp.~2198--2231, 2018.
	
	\bibitem{Wohlrab2018REEN}
	R.~Wohlrab, E.~Knauss, J.-P. Stegh{\"o}fer, S.~Maro, A.~Anjorin, and
	P.~Pelliccione, ``Collaborative traceability management: a multiple case
	study from the perspectives of organization, process, and culture,'' {\em
		Requirements Engineering}, vol.~25, pp.~21--45, 2020.
	
	\bibitem{Ebert2017}
	C.~Ebert and J.~Favaro, ``Automotive software,'' {\em IEEE Software}, vol.~34,
	pp.~33--39, May 2017.
	
	\bibitem{Wohlrab2019JSME}
	R.~Wohlrab, P.~Pelliccione, E.~Knauss, and M.~Larsson, ``Boundary objects and
	their use in agile systems engineering organizations,'' {\em Journal of
		Software: Evolution and Process}, vol.~31, no.~5, p.~e2166, 2019.
	
	\bibitem{Star1989}
	S.~L. Star and J.~R. Griesemer, ``Institutional ecology, `translations' and
	boundary objects: Amateurs and professionals in berkeley's museum of
	vertebrate zoology, 1907-39,'' {\em Social Studies of Science}, vol.~19,
	no.~3, pp.~387--420, 1989.
	
	\bibitem{Bond2003}
	J.~V. {Bond, III}, G.~H. Engelman, J.~Ekmark, J.~L. Jansson, M.~N. Tarabishy,
	and L.~Tellis, ``Collision mitigation by braking system,'' Aug. 2003.
	\newblock US Patent 6,607,255 B2.
	
	\bibitem{Bjarnason2017}
	E.~Bjarnason and H.~Sharp, ``The role of distances in requirements
	communication: a case study,'' {\em Requirements Engineering}, vol.~22,
	pp.~1--26, Mar 2017.
	
	\bibitem{Kim1998}
	D.~H. Kim, ``The link between individual and organizational learning,'' {\em
		The strategic management of intellectual capital}, vol.~41, p.~62, 1998.
	
	\bibitem{Mirarab2007}
	S.~Mirarab, A.~Hassouna, and L.~Tahvildari, ``Using bayesian belief networks to
	predict change propagation in software systems,'' in {\em ICPC'07},
	pp.~177--188, 2007.
	
\end{thebibliography}
\end{document}